\documentclass[preprintnumbers,floatfix,letterpaper,aps,prd,nofootinbib,superscriptaddress,onecolumn]{revtex4-2}

\usepackage{psfrag}
\usepackage{mathrsfs}
\usepackage{amssymb, bm}
\usepackage{amsmath, amsthm}
\usepackage{epstopdf}
\usepackage{hyperref}
\usepackage{enumerate}
\usepackage{longtable}
\usepackage{subfigure}
\usepackage{color}
\usepackage{mathrsfs}
\usepackage{graphicx}
\usepackage{bm,natbib,url,textcase}
\usepackage[english]{babel}
\usepackage[T1]{fontenc}
\usepackage{comment}
\usepackage{accents}

\allowdisplaybreaks

\hypersetup{
 colorlinks=true, % false: boxed links; true: colored links
 linkcolor=blue, % the colour of internal links
 citecolor=magenta, % the colour of links to the bibliography
 urlcolor=cyan % colour of external links
}

\definecolor{pinegreen}{rgb}{0.0, 0.47, 0.44}

\usepackage{tikz}

\definecolor{purple}{rgb}{1,0,1}
\definecolor{lime}{HTML}{a6CE39} % needs xcolor

\newcommand{\orcidicon}{%
	\begin{tikzpicture}
		\draw[lime, fill=lime] (0,0) 
		circle [radius=0.15] 
		node[white] {{\fontfamily{qag}\selectfont \tiny ID}};
		\draw[white, fill=white] (-0.0625,0.095) 
		circle [radius=0.007];
	\end{tikzpicture}	\hspace{-2mm}
}

\newcommand\orcidMarcello{{\href{https://orcid.org/0000-0003-0397-2705}{\orcidicon}}}
\newcommand\orcidSerena{{\href{https://orcid.org/0000-0002-8094-0865}{\orcidicon}}}
\newcommand\orcidAndrea{{\href{https://orcid.org/0000-0003-0329-2726}{\orcidicon}}}
\newcommand\orcidLavinia{{\href{https://orcid.org/0000-0002-5766-8242}{\orcidicon}}}

\def\nn{\nonumber}

\begin{document}

\title{First-order thermodynamics of Horndeski cosmology}

\author{Marcello Miranda \orcidMarcello}
\email{marcello.miranda@unina.it}
\affiliation{Scuola Superiore Meridionale, Largo San Marcellino 10, 
I-80138, Napoli, Italy}
\affiliation{Istituto Nazionale di Fisica Nucleare, Sez. di Napoli, Compl. Univ. Monte S. Angelo, 
Edificio G, Via	Cinthia, I-80126, Napoli, Italy}
\author{Serena Giardino \orcidSerena}
\email{serena.giardino@aei.mpg.de}
\affiliation{Max Planck Institute for Gravitational Physics (Albert 
Einstein Institute), Am Mühlenberg 1, 14476 Potsdam, Germany}
\affiliation{Institute for Theoretical Physics, Heidelberg 
University, Philosophenweg 16, 69120 Heidelberg, Germany}
\author{Andrea Giusti \orcidAndrea}
\email{agiusti@phys.ethz.ch}
\affiliation{Institute for Theoretical Physics, ETH Zurich, 
Wolfgang-Pauli-Strasse 27, 8093 Zurich, Switzerland}
\author{Lavinia Heisenberg \orcidLavinia}
\email{l.heisenberg@thpyhs.uni-heidelberg.de}
\affiliation{Institute for Theoretical Physics, Heidelberg 
University, Philosophenweg 16, 69120 Heidelberg, Germany}

\begin{abstract}

We delve into the first-order thermodynamics of Horndeski gravity, focusing on spatially flat, homogeneous, and isotropic cosmologies. Our exploration begins with a comprehensive review of the effective fluid representation within viable Horndeski gravity. Notably, we uncover a surprising alignment between the constitutive relations governing the ``Horndeski fluid'' and those of Eckart's thermodynamics.
Narrowing our focus, we specialize our discussion to spatially flat Friedmann-Lema{\^i}tre-Robertson-Walker spacetimes. Within this specific cosmological framework, we systematically analyze two classes of theories: shift-symmetric and asymptotically shift-symmetric. These theories are characterized by a non-vanishing braiding parameter, adding a nuanced dimension to our investigation.
On the one hand, unlike the case of the ``traditional'' scalar-tensor gravity, these peculiar subclasses of viable Horndeski gravity never relax to General Relativity (seen within this formalism as an equilibrium state at zero temperature), but give rise to additional equilibrium states with non-vanishing viscosity. On the other hand, this analysis further confirms previous findings according to which curvature singularities are ``hot'' and exhibit a diverging temperature, which suggests that deviations of scalar-tensor theories from General Relativity become extreme at spacetime singularities.
Furthermore, we provide a novel exact cosmological solution for an asymptotically shift-symmetric theory as a toy model for our thermodynamic analysis. 
\end{abstract}

\date{\today}

\maketitle

\section{Introduction}
\label{sec:1}
Scalar fields are of fundamental importance in cosmology, since they are used to address various puzzles in our understanding of the cosmic evolution, from early to late times. In the early universe, a minimally coupled scalar field rolling down a potential is the essence of the inflationary mechanism, although scalars are also widely employed in alternative scenarios such as genesis and bouncing cosmologies \cite{Brandenberger:2016vhg}. At late times, modified gravity theories adding a scalar field to the tensorial degrees of freedom of General Relativity (GR), and quintessence, namely, a canonical scalar field endowed with a potential, are among the most promising alternatives to a fine-tuned cosmological constant as dark energy \cite{Amendola:2015ksp}. Other than their versatility, a practical reason for this ubiquity of scalar fields in cosmological settings is that they can yield accelerated expansion without breaking isotropy, with a background field configuration $\phi=\phi(t)$. In order to be viable as dark energy candidates, scalars need to be very light, with a mass of the order of $m\simeq10^{-33}$ eV, so that modifications to GR would appear only on very large cosmological scales. At these scales, GR is not as well-tested as within the Solar System, and therefore there is still room for modifying our theory of gravity \cite{Heisenberg:2018vsk}.

The Horndeski class of scalar-tensor theories is the most general class exhibiting second-order equations of motion, irrespective of the specific background considered, therefore avoiding Ostrogradski instabilities \cite{Horndeski:1974wa, Kobayashi:2019hrl} (although some higher-order scalar-tensor theories beyond Horndeski, like DHOST, still admit second-order equations when a degeneracy condition is satisfied \cite{Langlois:2015cwa, Motohashi:2016ftl, Bernardo:2019vln}). Given their generality, Horndeski's theories encompass a plethora of scalar-tensor models that have been explored since the first attempt by Brans and Dicke \cite{BransDicke}, and all have impactful cosmological implications. A non-exhaustive list includes ``traditional'' scalar-tensor theories \cite{ST-1,ST-2,ST-3} (which contain $f(R)$ gravity as a subclass \cite{Sotiriou:2008rp,Capozziello:2011et}), quintessence \cite{Caldwell:1997ii}, k-essence \cite{Armendariz-Picon:2000nqq}, Galileon models endowed with shift and Galilean symmetries \cite{Kobayashi:2010cm,DeFelice:2010pv} and even a proxy theory to massive gravity \cite{deRham:2011by,Heisenberg:2014kea}.
Interesting cosmological consequences of such scalar-tensor theories include, for example, de Sitter attractors for shift-symmetric Lagrangians \cite{Kobayashi:2010cm}, the presence of scaling solutions in some Horndeski subclasses \cite{Gomes:2015dhl}, and multi-faceted applications of the Galileons, such as inflation \cite{Burrage:2010cu} and dark energy \cite{Gannouji:2010au}. 
Given the rich landscape of cosmological implications of Horndeski theories, any approach to such models finds its natural arena in a Friedmann-Lema\^{i}tre-Robertson-Walker (FLRW) spacetime.

%%%%%%%%%%%thermodynamics and outline
A recent formalism with intriguing applications to Horndeski theories is the so-called first-order thermodynamics of modified gravity, devised in \cite{Faraoni:2021lfc, Faraoni:2021jri} and briefly reviewed in \cite{Giardino:2023ygc}, which provides a concrete realisation of the ideas in \cite{jacobson, Eling:2006aw}, albeit in a different setting. Its goal is the construction of a unifying perspective on the landscape of gravity theories, comprised of GR and its generalisations. The essence of first-order thermodynamics, originally conceived for ``traditional'' scalar-tensor theories \cite{Faraoni:2021jri}, involves isolating the contribution of the scalar $\phi$ to the effective stress-energy tensor in the Einstein equations, which is known to take the form of an imperfect fluid \cite{Faraoni:2018qdr, Nucamendi:2019uen, Pimentel:1989bm, Miranda:2022uyk}. The novelty of the formalism comes in when we apply Eckart's non-equilibrium thermodynamics to this fluid, which entails first-order constitutive relations in the dissipative variables. This leads to the identification of the fluid's effective temperature, a sort of ``temperature of scalar-tensor gravity'', which is nothing but a temperature relative to GR (the equilibrium state at zero temperature).
This effective temperature is positive definite for theories containing an additional scalar degree of freedom with respect to GR, characterising these scalar-tensor theories as non-equilibrium states, in a sort of thermodynamics of gravitational theories. Moreover, the temperature is the order parameter ruling the dissipative approach to equilibrium, described by an effective heat equation, which often entails a relaxation to the GR equilibrium, especially in cosmological settings \cite{Giardino:2022sdv}. 

The formalism has been extended to various situations \cite{Giardino:2022sdv, Giardino:2023qlu, Faraoni:2022doe, Faraoni:2022gry}, but it was for Horndeski theories that it showed the most interesting consequences \cite{Giusti:2021sku}. Namely, the thermodynamical analogy described above irreparably breaks down for the most general Horndeski theories, and only holds in the ``viable'' Horndeski subclass that predicts gravitational-wave propagation at the speed of light. This connects the formalism, so far purely theoretical, with the observational constraints placed on Horndeski theories by the multi-messenger event GW170817/GRB170817A \cite{LIGOScientific:2017ync, LIGOScientific:2017zic}.

Motivated by this development, the goal of the present work is to extend the first-order thermodynamics of Horndeski theories \cite{Giusti:2021sku} to the fruitful setting of FLRW spacetime, in order to test the physical intuition provided by the formalism.
The paper is organized as follows: in Section \ref{sec:2}, we review Horndeski theories and the effective fluid approach, which makes it possible to formulate a thermodynamical description. In Section \ref{sec:3} we discuss the salient features of the first-order thermodynamics of viable Horndeski and specify to a cosmological background to explore its physical implications. In Section \ref{sec:4} we apply the formalism to some exact cosmological solutions of viable Horndeski gravity (or subclasses thereof), which exhibit particularly intriguing properties.

\section{Horndeski theories and effective fluid approach}
\label{sec:2}

%%%%%%%%%horndeski lagrangian
The full Horndeski action is given by 
\begin{equation}
S\left[ g_{ab}, \phi \right] = \frac{1}{2}\int d^4 x \sqrt{-g} \, \left( {\cal L}_2 + 
{\cal L}_3+ {\cal L}_4+ 
{\cal L}_5 \right) + S^\mathrm{(m)} \,, \label{Horndeskiaction}
\end{equation}
where
\begin{align}
{\cal L}_2 =\,& G_2\left( \phi, X \right) \,,\\
%&\nn\\
{\cal L}_3 =\,&- G_3\left( \phi, X \right) \Box \phi \,,\\ 
%&\nn\\
{\cal L}_4 =\,& G_4\left( \phi, X \right) R +G_{4X} \left( \phi, X \right) 
\left[ \left( \Box \phi \right)^2 -\left( \nabla \nabla \phi \right)^2 
\right] \,, \\ 
%&\nn\\
{\cal L}_5  =\,&  G_5\left( \phi, X \right) G_{ab}\nabla^a \nabla^b \phi 
-\frac{ G_{5X} }{6} \left[ \left( \Box\phi \right)^3 -3\Box\phi 
\left( \nabla \nabla \phi \right)^2 +2\left( \nabla \nabla \phi 
\right)^3 \right] \,.
\end{align}
With $\nabla_a$ we indicate the covariant derivative associated to the Levi-Civita connection of the metric, $G_{ab}=R_{ab}-\dfrac{1}{2}Rg_{ab}$ is the Einstein tensor, with $R_{ab}$ and $R$ the Ricci tensor and the Ricci scalar, respectively, $S^{(m)}$ is the matter action. The functions $G_i(\phi,X)$ $(i=2,3,4,5)$ are arbitrary regular functions of the theory, where $X\equiv-\dfrac{1}{2}\nabla_a\phi\nabla^a\phi$ is the canonical kinetic term of the scalar $\phi$. Their partial derivatives are denoted as $G_{i\phi}\equiv\partial G_i/\partial\phi$ and $G_{iX}\equiv\partial G_i/\partial X$. Note the compact notation adopted to indicate $ \left( \nabla \nabla \phi \right)^2 
\equiv \nabla_a\nabla_b\phi\nabla^a \nabla^b \phi$ and $ \left( 
\nabla\nabla \phi \right)^3 \equiv \nabla_a\nabla_c\phi 
\nabla^c\nabla^d \phi \nabla_d \nabla^a \phi $. Throughout this work, $8\pi G=c=\hslash=1$ and the metric signature is $(-+++)$.

%%%%%%%%%%%%viable horndeski
The multi-messenger event GW170817/GRB170817A \cite{LIGOScientific:2017ync} confirmed with remarkable precision that gravitational waves travel at the speed of light, therefore imposing strong constraints on those sectors of Horndeski theories that do not fulfil this requirement \cite{Baker:2017hug, Creminelli:2017sry, Ezquiaga:2017ekz, Heisenberg:2018vsk}. The class of viable Horndeski theories that exactly respects this constraint is characterised by \footnote{ Note that the LIGO/Virgo constraint on the speed of gravitational waves is restricted to frequencies $10-100$ Hz. This is at the edge of the strong coupling scale of Horndeski
theories, where the regime of validity of the effective field theory breaks down, and, potentially, new operators at this cutoff scale could affect the propagation speed \cite{deRham:2018red}.}
\begin{equation}
G_{4X}=G_5=0.
\label{viable}
\end{equation}
In the following, we will focus on such viable Horndeski theories, where first-order thermodynamics holds, which also greatly simplifies the analysis.
%
%
%\newpage
%
%
\subsection{Effective stress-energy tensor}
Performing the variation of the Action~\eqref{Horndeskiaction} with respect to the metric tensor $g^{ab}$ and the scalar field $\phi$, we obtain the corresponding field equations, 
\begin{eqnarray}
 G_4  \, G_{ab} -\nabla_{a}\nabla_{b}G_4  
+ \left[ \Box G_4 -\dfrac{G_2  }{2} 
-\dfrac{1}{2} \, \nabla_{c} 
\phi\nabla^{c}G_{3} 
\right] g_{ab} %&&\nonumber\\
 + \frac{1}{2}\left[ G_{3X} \, 
\Box\phi -G_{2X} \right] 
\nabla_{a}\phi\nabla_{b}\phi 
+ \nabla_{(a}\phi \nabla_{b)}G_{3} %&&\,
=  T^\mathrm{(m)}_{ab} \,,\label{hfeq}
\end{eqnarray}
and
\begin{eqnarray}
G_{4\phi}  R + G_{2\phi} +G_{2X}  
\Box\phi+\nabla_{c}\phi\nabla^{c}G_{2X}%&&\vspace{7pt}\nn
%&&\nn\\
-G_{3X} (\Box\phi)^2-\nabla_{c}\phi\nabla^c 
G_{3X}  \Box\phi-G_{3X} \nabla^{c} 
\phi\Box\nabla_c\phi\vspace{7pt}%&&\nn\\\
&&\nn\\
+G_{3X}  R_{ab}\nabla^{a} 
\phi\nabla^{b}\phi-\Box G_{3} -G_{3\phi} \Box\phi&&\,=0 \label{heom}\,,
\end{eqnarray}
where $T^{\rm (m)}_{ab}\equiv-\dfrac{2}{\sqrt{-g}}\dfrac{\delta S^{\rm(m)}}{\delta g^{ab}}$ is the matter stress-energy tensor. The presence of indices encompassed by parentheses indicates the symmetrization of the indices, while square brackets indicate the anti-symmetrization, defined as $V_{(ab)}=\dfrac{1}{2}(V_{ab}+V_{ba})$ and $V_{[ab]}=\dfrac{1}{2}(V_{ab}-V_{ba})$, respectively. 

The Horndeski field equations~\eqref{hfeq} can be recast in the form of Einstein equations,
\begin{equation}\label{Einstein}
    G_{ab}=  T^{(\rm eff)}_{ab}\,,
\end{equation}
where
\begin{equation}
    T^{(\rm eff)}_{ab}=\frac{T^{(\rm m)}_{ab}}{G_4}+T^{( \phi)}_{ab}\,, 
\label{teff}
\end{equation}
\begin{equation}
    T^{(\phi)}_{ab}=T^{(2)}_{ab}+T^{(3)}_{ab}+T^{(4)}_{ab},
    \label{tphi}
\end{equation}
and the individual contributions are
\begin{align}
      T^{(2)}_{ab}=\,&\frac{1}{2G_4} \left( G_{2X}\nabla_{a}\phi\nabla_{b}\phi+G_{2}\,g_{ab} \right)\,,\\
    &\nn\\
       T^{(3)}_{ab}=&\frac{1}{2G_4} \left( G_{3X} \nabla_{c} X  \nabla^{c} \phi  - 2 X G_{3\phi} \right) g_{ab} \nn\\
    \,& - \frac{1}{2G_4}\left( 2 G_{3\phi} + G_{3X} \Box \phi \right) \nabla _a  \phi \nabla _b \phi- \frac{G_{3X}}{G_4} \nabla_{(a} X \nabla_{b)} \phi\,,\\
    &\nn\\
       T^{(4)}_{ab}=\,&\frac{G_{4\phi}}{G_4}(\nabla_{a}\nabla_{b}\phi-g_{ab}\Box\phi)+\frac{G_{4\phi\phi}}{G_4}(\nabla_{a}\phi\nabla_{b}\phi+2X\,g_{ab})\,.
\end{align}
The equation of motion for the scalar field can be written as
\begin{equation}
    \mathcal{S}_2+\mathcal{S}_3+\mathcal{S}_4=0\,,
    \label{phieom}
\end{equation}
where
\begin{align}
    \mathcal{S}_2=&\left(G_{2X}g^{ab}-G_{2XX}\nabla^{a}\phi\nabla^{b}\phi\right)\nabla_{a}\nabla_{b}\phi+G_{2\phi}-2XG_{2\phi X}\,,\\
    &\nn\\
    \mathcal{S}_3=&\,G_{3X}R_{ab}\nabla^{a}\phi\nabla^{b}\phi-2\left(G_{3X}g^{ab}g^{cd}-G_{3XX}\nabla^{a}\phi\nabla^{b}\phi \,g^{cd}\right)\nabla_{[a|}\nabla_{b}\phi\nabla_{|c]}\nabla_{d}\phi\nn\\
    &-2\left[\left(G_{3\phi}-XG_{3\phi X}\right)g^{ab}-G_{3\phi X}\nabla^{a}\phi\nabla^{b}\phi\right]\nabla_{a}\nabla_{b}\phi\,+2XG_{3XX},\\
    &\nn\\
     \mathcal{S}_4=&\,G_{4\phi}R\,.
\end{align}

It is well known (see Ref.~\cite{Faraoni:2018qdr}) that the scalar contribution $T_{ab}^{(\phi)}$ to the effective stress-energy tensor $T_{ab}^{(\rm eff)}$ can be recast in the imperfect fluid form 
\begin{align}
  T^{ab}%&=\rho^{(\phi)} u_{a}u_{b}+q_{a}^{(\phi)}u_{b} + q_{b}^{(\phi)} u_{a} +\Pi^{(\phi)}_{ab},
  %\nn\\
  &=\rho u^{a}u^{b}+Ph^{ab}+2q^{(a}u^{b)} +\pi^{ab}\,,
  \label{imperfect}
\end{align}
where the 4-vector $u^{a}$ is the fluid's 4-velocity ($u_{a}u^{a}=-1$), 
$h_{ab} = g_{ab}+u_{a} u_{b}$ is the projector onto the 
3-space orthogonal to $u^c$, $\rho = T^{ab}u_{a}u_{b}$ is the energy density, 
$P=\tfrac{1}{3}\,T^{ab}h_{ab}$ is the isotropic pressure, $q^{a}=-T^{cd}u_{c}h_{d}{}^{a}$ 
is the heat flux density, $\pi^{ab} = \left(h^{ac} h^{bd} -\frac{1}{3}h^{cd}h^{ab} \right)T_{cd}$ is 
the traceless part of the stress tensor, describing the anisotropic stresses.
%
\begin{comment}
    with $q_a^{(\phi)}$ the heat flux density, $\rho^{(\phi)}$ the energy density, $\Pi_{ab}^{(\phi)}=P^{(\phi)}h_{ab}+\pi^{(\phi)}_{ab}$ the stress tensor containing the isotropic pressure $P^{(\phi)}$ and the anisotropic stress tensor $\pi_{ab}^{(\phi)}$.
\end{comment}
%
Assuming that the scalar field gradient is timelike, 
$\nabla_{a}\phi\nabla^{a}\phi<0$, it is possible to define the 4-velocity of the effective fluid as follows
\begin{equation}
u^{a} \equiv\epsilon\frac{\nabla^{a}\phi}{\sqrt{2X}}\,,
\end{equation}
where $\epsilon=\pm 1$ is used to ensure a future-oriented 4-velocity. Then, the derivatives on the scalar field can be written as 
\begin{gather}
    \nabla_{a}\phi=\,\epsilon\sqrt{2X}\,u_{a}\,,\qquad    \nabla_{a}X=\,-\dot{X}\,u_{a}-2X\,\dot{u}_{a}\,,\label{fstD}\\
    \nn\\
    \nabla_{a}\nabla_{b}\phi=\,\epsilon\sqrt{2X} \left( \nabla_{a}u _{b}-\dot{u}_{a}u_{b}\right)-\epsilon\frac{\dot{X}}{\sqrt{2X }} \, u _{a}u _{b}\,,\label{ddf}\qquad \Box\phi=\,\epsilon\left(\sqrt{2X} \, \Theta+\frac{\dot{X} }{\sqrt{2X }}\right) ,
\end{gather}
where $\Theta\equiv\nabla_a u^a$ is the expansion scalar and $\dot{u}^a\equiv u^{b}\nabla_{b}u^{a}$ is the 4-acceleration of the fluid.

In the following, we adopt the decomposition $\nabla_{b}u_{a}=\sigma_{ab}+\frac{1}{3}\Theta\,h_{ab}+\omega_{ab}-\dot{u}_{a}u_{b}$, where $\sigma^{ab} \equiv \left(h^{ac} h^{bd} -\frac{1}{3}h^{cd}h^{ab} \right) \nabla_{(c} u_{d)}$  is the shear tensor and $\omega^{ab} \equiv h^{ac} h^{bd} \nabla_{[d} u_{c]}$ is the vorticity tensor. The latter vanishes since the effective fluid is derived from the scalar field gradient. One can easily check it from the torsionless property of the covariant derivative,
    \begin{equation}
        \nabla_{a}\nabla_{b}\phi=\nabla_{b}\nabla_{a}\phi\quad\Rightarrow\quad\omega_{ba}=\omega_{ab}\,,
    \end{equation}
which implies that the vorticity tensor vanishes identically because of the antisymmetry of $\omega_{ab}$.

The Eq.~(\ref{fstD}.b) is obtained by rewriting $\nabla_{a}=h_{a}{}^{b}\nabla_{b}-u_{a} \, u^{b}\nabla_{b}$ and using the following relation
\begin{align}\label{grdX}
    h_{ab}\nabla^{b}X=\,&-\nabla_{b}\phi\nabla^b\nabla_a\phi-\frac{1}{2X}\nabla_{b}\phi\nabla_{a}\phi\nabla_{c}\phi\nabla^{b}\nabla^{c}\phi\nn\\
    =\,&-2X\dot{u}_a\,.
\end{align}

In this framework, the effective stress-energy tensor of the $\phi$-fluid reads
\begin{align}\label{tphiimp}
      T^{(\phi)}_{ab} =\, & \Bigg[ \frac{2XG_{2X}-G_{2} -2XG_{3\phi}}{2G_4}+\epsilon\frac{\sqrt{2X}  \left( G_{4\phi} -  X G_{3X} \right)}{  
    G_{4}}\,\Theta \Bigg]u_{a} u_{b}\nn\\
    &+ \Bigg[\frac{ G_{2} + 4 X G_{4\phi\phi}  -2 X G_{3\phi}}{2  
    G_4 } -  \epsilon\frac{ \left( G_{4\phi} - X G_{3X} 
    \right)}{\sqrt{2X}\,  G_{4} 
    }\,\dot{X} \,-\epsilon\frac{2 \sqrt{2X} \,  G_{4\phi} \Theta}{3  G_4 }\, \Bigg] h_{ab}\nn\\
    &\nn\\
    &-\epsilon\frac{2\sqrt{2X} \left( G_{4\phi} - X G_{3X} \right)}{  G_{4}}\dot{u}_{(a} u_{b)}+\epsilon\frac{\sqrt{2X} \, G_{4\phi} }{ G_{4}} \, \sigma _{ab}\,.
\end{align}

Comparing Eq.~\eqref{tphiimp} with the generic imperfect fluid stress-energy tensor in Eq.~\eqref{imperfect}, we can now extract the characteristic quantities of the scalar effective fluid: 
\begin{align}
   \rho^{(\phi)}=\, & \frac{1}{2G_4}\left( 2XG_{2X}-G_{2} -2XG_{3\phi} \right)+\epsilon\frac{\sqrt{2X}}{G_4} \left( G_{4\phi}-XG_{3X} \right)\Theta\,,\label{rho}\\
  &\nn\\
    P^{(\phi)}=\,&\frac{1}{2G_4} \left( G_2-2XG_{3\phi}+4XG_{4\phi\phi} \right)-\epsilon\frac{\left(G_{4\phi}-XG_{3X}\right)}{G_4\sqrt{2X}}\dot{X}-\epsilon\frac{2G_{4\phi}}{3G_4}\sqrt{2X}\,\Theta\,\nonumber\\
  =\,&\frac{1}{2G_4} \left( G_2-2XG_{3\phi}+4XG_{4\phi\phi} \right)-\frac{\left(G_{4\phi}-XG_{3X}\right)}{G_4}\,\Box{\phi}+\epsilon\frac{\left(G_{4\phi}-3XG_{3X}\right)}{3G_4} \, \sqrt{2X}\,\Theta\,,\label{P}\\
  &\nn\\
    q_{a}^{(\phi)}=\,&-\epsilon\frac{\sqrt{2X} \left( G_{4\phi} - X G_{3X} \right)}{  G_{4}}\dot{u}_{a}\,,\label{q}\\
  &\nn\\
    \pi_{ab}^{(\phi)}=\,&\epsilon\frac{\sqrt{2X} \, G_{4\phi} }{ G_{4}} \, \sigma _{ab}\,.\label{pi}
\end{align}
This formal rewriting of $T^{(\phi)}_{ab}$ takes a deeper meaning in the context of dissipative fluids. Such fluids are classified according to their constitutive relations. 
As compellingly shown in \cite{Pujolas:2011he}, exploring the properties of the imperfect fluid behind modified theories of gravity allows one to obtain an intuitive picture of their physical meaning, often obfuscated by cumbersome expressions. The effective fluid approach provides a promising way to classify different Horndeski subclasses based on the nature of this fluid. In particular, in \cite{Miranda:2022wkz}, the requirement that such a fluid be Newtonian (i.e., with the viscous stresses depending only on the first derivatives of the
fluid’s 4-velocity) was explored. This requirement is relevant since, as will become clear in the following section, the simplest non-equilibrium thermodynamical treatment that we are interested in restricts to first-order derivatives in the fluid quantities. In order to understand the dissipative properties of the effective scalar fluid we are dealing with, we need to write the derivatives of the scalar field in terms of 4-velocity gradients. The only problem in this task arises when considering the pressure. Indeed, inside Eq.~\eqref{P} a $\Box\phi$-contribution is present (or, equivalently, a term containing $\dot{X}$, because of the linearity of Eq.~\eqref{fstD}). Therefore, the only way to completely translate $\Box\phi$ into the effective fluid formalism is by taking into account the equation of motion of the scalar field~\eqref{phieom}.

Using the metric field equation~\eqref{Einstein} to rewrite the curvature contributions inside Eq.~\eqref{phieom} in terms of the total effective stress-energy tensor~\eqref{teff} through $R=-  T^{(\rm eff)}$ and $R_{ab}= \left(T^{\rm (eff)}_{ab}-\frac{1}{2}T^{(\rm eff)}g_{ab}\right)$, it is possible to algebraically solve the scalar field equation of motion and obtain $\Box\phi$. This yields
\begin{equation}\label{boxf}
\Box\phi=\frac{A+B\,\Theta+C\Theta^2+D\,\sigma_{ab}\sigma^{ab}+E\,\dot{u}^c\dot{u}_c}{J+K\,\Theta},   
\end{equation}
where
\begin{align}
    A=&\,  T^{\rm(m)} (G_{4\phi}  - X G_{3X}) -2 \left(T^{\rm(m)}_{ab} u^{a} u^{b}\right) X G_{3X}+G_{2} (2 G_{4\phi} - X G_{3X}) - G_{4}G_{2\phi}\nn\\
     &X \left[(G_{4\phi}-X G_{3X}) (6 G_{4\phi\phi} + G_{2X}  -4 G_{3\phi}) +2 G_{4\phi} G_{3\phi} + 2 G_{4}( G_{2\phi X} -G_{3\phi\phi})\right]\,,\\
     &\nn\\
    B=&\,\epsilon(2X)^{3/2} \left[2 X G_{3X}^2-2 G_{4\phi} G_{3X} + G_{4}(G_{2XX}  -2 G_{3\phi X})\right]\,,\\
    &\nn\\
    C=&\,- \frac{4}{3} X G_{4} (2 G_{3X} + 3 X G_{3XX})\,,\\
    &\nn\\
    D=&\,-2 X G_{4}G_{3X}\,,\\
    &\nn\\
    E=&\,4 X G_{4} (G_{3X} + X G_{3XX})\,,\\
    &\nn\\
    J=&\,3(G_{4\phi}-X G_{3X})^2 + G_{4} \left[G_{2X} + 2 X G_{2XX}  -2 (G_{3\phi}+ X G_{3\phi X})\right]\,,\label{eq:J}\\
    &\nn\\
    K=&\,-2 \epsilon\sqrt{2X} G_{4}(G_{3X} + X G_{3XX})\,.\label{eq:K}
\end{align}
Eq.~\eqref{boxf} casts the $\Box\phi$ in terms of the kinematic quantities of the effective fluid identified above. It is of course valid as long as the denominator $J+K\Theta$ is not vanishing\footnote{{ Here we are just treating Eq.~\eqref{boxf} as an algebraic equation for $\Box\phi$, intending to rewrite it in terms of the kinematic quantities.}}. In particular, theories for which the denominator of Eq.~\eqref{boxf} identically vanishes correspond to the non-dynamical class of Horndeski theories, which includes the extended cuscuton model \cite{Iyonaga:2018vnu}.
Indeed, $J+K\Theta=0$ entails $K=0$ and $J=0$, separately \cite{Miranda:2022wkz}. On the one hand, given Eq.~\eqref{eq:K}, $K=0$ implies $G_{3X} + X G_{3XX}=0$, which has
\begin{equation}
    G_3(\phi,X)=F(\phi)\ln( X/X_{*})+V(\phi)\,,
    \label{G3_EC}
\end{equation}
as a general solution, with $X_{*}$ constant. On the other hand, $J=0$ provides the functional form of $G_2$,
\begin{equation}
    G_2(\phi ,X)= \mu(\phi)\sqrt{2X}+\nu(\phi )-4X \left( F_{\phi}(\phi)+ \frac{3\left[F(\phi)-G_{4\phi}(\phi)\right]^2}{4G_4(\phi )}+\frac{1}{2} V_{\phi}(\phi)\right)+2 F_{\phi}(\phi)X \ln (X/X_{*})\,\label{G2_EC},
\end{equation}
where $F(\phi)$, $\mu(\phi)$ and $\nu(\phi)$ are generic functions.
It is straightforward to verify that the potential $V(\phi)$ does not play any role since it can be eliminated by performing an integration by parts, namely $-V(\phi) \, \Box\phi\simeq2 X\,V_{\phi}(\phi) $ up to a total derivative. Therefore, $G_{3}=V(\phi)$ is equivalent to considering $\tilde{G}_3=0$ and $\tilde{G}_2=G_{2}+2XV_{\phi}$.
Redefining $F\to G_{4\phi}+\tfrac{1}{2}F$, Eq.~\eqref{G2_EC} (and Eq.~\eqref{G3} in the following) turn into the well-known form used in \cite{Iyonaga:2018vnu, Iyonaga:2020bmm, Miranda:2022brj}.

{It is worth stressing that we need Eq.~\eqref{boxf} only in the case of $G_{4\phi}\neq X\,G_{3X}$. Indeed, when $G_{4\phi}=X\,G_{3X}$ (\textit{i.e.}, $F=G_{4\phi}$), $\Box\phi$ disappears from Eq.~\eqref{P} thus making Eq.~\eqref{boxf} no longer necessary for carrying out the thermodynamic analogy, and the fluid behaves as a Newtonian fluid~\cite{Miranda:2022wkz}. An example of such a scenario is given by $k$-essence, for which one has $G_{4\phi}=G_{3X}=0$.}

After substituting Eq.~\eqref{boxf} into Eq.~\eqref{P} to obtain an expression for the pressure, we can also rewrite Eqs.~\eqref{rho},~\eqref{q} and~\eqref{pi} in a compact way, making the dependence on the 4-velocity gradients apparent:
\begin{align}
      \rho^{(\phi)}&=\,\rho_{0}-\xi\Theta\,,\label{rho_cr}\\
      P^{(\phi)}&=P_0+\xi\left(\frac{A+B\Theta+C \Theta^2+D \sigma^2+E\dot{u}^2}{J'+K' \Theta}\right)-\left(\xi-\frac{4}{3}\eta\right)\Theta\,,\label{P_cr}\\
      q^{(\phi)}_{a}&=\xi\dot{u}_{a}\,,\label{q_cr}\\
      \pi^{(\phi)}_{ab}&=-2\eta\sigma_{ab}\,,\label{pi_cr}
\end{align}
where $J'=\epsilon\sqrt{2X}J$, $K'=\epsilon\sqrt{2X}K$, $\rho_0=\left( 2XG_{2X}-G_{2} -2XG_{3\phi} \right)/{2G_4}$, $P_0=\left( G_2-2XG_{3\phi}+4XG_{4\phi\phi} \right)/{2G_4}$, $\xi=-\epsilon\sqrt{2X} \left( G_{4\phi} - X G_{3X} \right)/  G_{4}$, $\eta=-\sqrt{2X}G_{4\phi}/2G_{4}$, $\sigma^2=\sigma_{ab}\sigma^{ab}$, and $\dot{u}^{2}=\dot{u}_{c}\dot{u}^{c}$.

Note that this is still a formal rewriting, and only in the next section it will be connected to a dissipative thermodynamical description that provides the physical interpretation behind the coefficients. 

The viable Horndeski effective fluid is then characterized by linear constitutive relations for the energy density~\eqref{rho_cr}, the heat flux density~\eqref{q_cr}, and the anisotropic stress~\eqref{pi_cr}. The non-Newtonian behaviour of the fluid arises from the pressure~\eqref{P_cr}. The requirement of a Newtonian fluid is quite stringent and selects two specific subclasses of viable Horndeski: one is characterized by $G_{3}=G_{4\phi}\ln(X/X_{*})$ (associated to $\xi=0$), and the other is identified with $G_{3}=0$ \cite{Miranda:2022wkz}. This way, all the non-linear contributions in the dissipative quantities due to the presence of $\Box\phi$ in Eq.~\eqref{P_cr} disappear. These classes are disconnected with respect to conformal transformations of the metric tensor, and the second one exists only for a dynamical scalar field. More general theories correspond to effective fluids that are non-Newtonian, and therefore exotic and less easily interpretable from the physical point of view.

However, here we are interested in applying Eckart's thermodynamics in the context of cosmology, {\it i.e.} with a particular fixed background.
For some particular geometries, it is possible to realise an Eckart-like effective fluid in a bigger subclass of viable Horndeski, containing the previous classes as sub-cases. That is the case of FLRW universes with a homogeneous scalar field.\\

\section{First-order thermodynamics of Horndeski theories}
\label{sec:3}
%%%const equations
The non-equilibrium thermodynamics developed by Eckart allows us to obtain a ``thermodynamics of gravity theories'', in which GR represents the equilibrium state and scalar-tensor theories are non-equilibrium states, providing a concrete realization of the ideas in \cite{jacobson, Eling:2006aw}. Eckart's thermodynamics is distilled in three 
constitutive relations\footnote{In 
 a more recent formulation of the first-order thermodynamics of real fluids~\cite{Bemfica:2017wps, Bemfica:2020zjp, Bemfica:2019knx}, linear viscous contributions are present also in the expression of the energy density, similarly to Eq.~\eqref{q_cr}. Notice that here we work in the Eckart (or particle) frame.} that connect the viscous pressure $P_\text{vis}$ 
with the fluid expansion $\Theta$, the heat current density $q^a$
with  the temperature ${\cal T}$, and the anisotropic stresses $\pi_{ab}$ with 
the  shear tensor $\sigma_{ab}$:
\begin{eqnarray}
P_{\rm tot}&=&P_{\rm non-visc}+P_{\rm visc}\,,\label{EckartPtot}\\
&&\nonumber\\
P_\text{visc} &=& -\zeta \, \Theta \,,\label{EckartPvis}\\
&&\nonumber\\
q_a &=& -{\cal K} \left( h_{ab} \nabla^b {\cal T} + {\cal T} \dot{u}_a 
\right)
\,, \label{Eckartq}\\
&&\nonumber\\
\pi_{ab} &=& - 2\eta \, \sigma_{ab} \,,\label{Eckartpi}
\end{eqnarray}
where ${\cal K}$, $\zeta$ and $\eta$ are the thermal conductivity, bulk viscosity and
shear viscosity, respectively, and we generally assume $h_{ab} \nabla^b {\cal T}=0$.
The temperature of Horndeski gravity (inextricably linked to the thermal conductivity) reads \cite{Giusti:2021sku}
\begin{equation}
\mathcal{K} \mathcal{T} = \epsilon\frac{\sqrt{2 X} (G_{4 \phi} - X G_{3 
X})}{G_{4}},
\label{kt}
\end{equation}
and reduces to GR equilibrium state characterized by $\mathcal{K} \mathcal{T}=0$ if $\phi=\rm const$.

The most interesting finding is that this formalism does not work for the most general Horndeski theories, because some terms in their field equations explicitly break the proportionality required by the constitutive equations \cite{Giusti:2021sku}. These terms are precisely those that violate the equality between the propagation speeds of gravitational and electromagnetic waves. Therefore, the validity of first-order thermodynamics seems to be related to the physical viability of Horndeski theories, which is a very intriguing result.
The breaking of the thermodynamic analogy is also interesting from the purely theoretical point of view: it happens for the operators which contain derivative nonminimal couplings and nonlinear contributions in the connection. This relates to the well-known and long-standing problem of separating matter and gravity degrees of freedom in a local description.

More specifically, in \cite{Giusti:2021sku} it is shown that, whenever we try to apply the thermodynamic formalism to theories beyond the viable class, the effective stress-energy tensor contains the term
\begin{equation}
T_{ab}^{(\phi)} \supset \alpha(\phi, X) \, R_{a c b d} \nabla^c \phi \nabla^d \phi \, ,
\end{equation}
where $\alpha(\phi, X)$ is a generic function. It is precisely the Riemann tensor $R_{a c b d}$ which ends up breaking the proportionality between the traceless shear tensor $\sigma_{ab}$ and the anisotropic stress tensor $\pi_{ab}$, and thus Eckart's constitutive equations~\eqref{EckartPvis},~\eqref{Eckartq} and~\eqref{Eckartpi} no longer hold.

 %%%%%%Fluid nature 

\subsection{First-order thermodynamics in FLRW background}
%%%%%%%%flrw cosmo

In \cite{Giardino:2022sdv}, the first-order thermodynamics of ``traditional'' scalar-tensor theories was studied in an FLRW background, with the goal of testing the physical intuition behind the formalism on some well-known exact solutions. The main result is that the GR equilibrium
state of zero temperature is almost always approached at late times $t\rightarrow +\infty$ throughout the cosmic expansion, while the behaviour expected for singularities (namely ${\cal KT}\rightarrow +\infty$, indicating an extreme deviation of the theory from GR equilibrium) is confirmed for solutions endowed with an initial singularity. Compellingly, this result about scalar-tensor theories ``relaxing'' to GR in a cosmological setting echoes those of \cite{Damour:1992kf, Damour:1993id}, albeit in a very different context. We are now in a position to perform the same feat as \cite{Giardino:2022sdv} with the more general class of viable Horndeski theories.

The FLRW line element reads 
\begin{equation}
    ds^2=-dt^2+a^2\left(\frac{dr^2}{1-kr^2}+r^2d\Omega^2\right),
\end{equation}
where, $a=a(t)$ is the scale factor of the FLRW universe, $k=0,\,\pm1$ is a parameter identifying the curvature of the 3-space, and $ d\Omega^2 \equiv d\vartheta^2 +\sin^2 \vartheta \, d\varphi^2$ is the line element on the unit 2-sphere. In particular, we restrict our discussion to the spatially flat case, {\it i.e.}, $k=0$.

The 4-velocity of the effective fluid in a FLRW setting becomes
\begin{equation}\label{4velocity}
 u^{a}\equiv\epsilon\frac{\nabla^{a}\phi}{\sqrt{2X}}=\left(-\epsilon\,{\rm Sign}(\dot{\phi}),0,0,0\right),
\end{equation}
where we assume that $\phi$ is strictly monotonic in $t$. Then, \eqref{4velocity} is future-oriented only if $\epsilon=-{\rm Sign}(\dot{\phi})$. As a consequence, the equation $\dot{\phi}=-\epsilon\sqrt{2X}$ holds, since $X=\frac{1}{2}\dot{\phi}^2$. 

As mentioned in the previous section, once we work with a fixed background, the constraint that an effective fluid is linear in $\nabla_{b}u_a$ is less stringent than in the general case with any geometry.
The features of the FLRW metric allow us to find a larger subclass of viable Horndeskis containing the previous classes as subcases. This is the case of FLRW universes with a homogeneous scalar field.

The expansion scalar in FLRW reads $\Theta=3H$, and the shear tensor and the 4-acceleration vanish ($\sigma^2=0$, $\dot{u}^2=0$).
Moreover, the Friedmann constraint reads
\begin{equation}\label{time_feqs}
    H^2=\frac{ 1 }{3}\left(\frac{\rho^{\rm(m)}}{G_4}+\rho^{(\phi)}\right)=\frac{ 1 }{3}\left(\frac{\rho^{\rm(m)}}{G_4}+\rho_0-3H\xi\right)\,.
\end{equation}
Therefore, since $\rho^{(\phi)}$ is always linear in $H$, {\it i.e.}, linear in the expansion scalar, we can rewrite the $\Theta^2=9H^2$ term in Eq.~\eqref{boxf} and Eq.~\eqref{P_cr} as a linear expression in terms of $\Theta=3H$. Then, the general expression for the pressure takes the form
\begin{equation}
    P^{(\phi)}=P_0+\xi\left(\frac{A'+B'\Theta}{J'+K' \Theta}\right)-\left(\xi-\frac{4}{3}\eta\right)\Theta\,.
\end{equation}
At this point, Eckart's constitutive relation~\eqref{EckartPvis} can be realised by imposing $K'=0$, which corresponds to assuming 
\begin{equation}\label{G3}
    G_{3}(\phi, X)=F(\phi)\ln(X/X_{*})\,.
\end{equation}
This functional form is a solution of the partial differential equation $G_{3X} + X G_{3XX}=0$ (see Eq.~\eqref{eq:K}), which eliminates the non-linear contribution due to the denominator in Eq.~\eqref{P_cr}. As mentioned above, an additional function of the scalar field, $V(\phi)$, in $G_3$ is neglected since it can be reabsorbed through a redefinition of $G_2$. Therefore, in order to deal with a linear effective fluid and apply the Eckart's thermodynamics, in the following we assume the above functional form of $G_{3}$. This particular choice is not just attractive for its simplicity, but also includes interesting applications like shift-symmetric theories exhibiting hairy black holes \cite{Fang:2018vog, Bernardo:2019vln}, vanishing braiding theories \cite{Noller:2020lav}, and the case $XG_{3X}\propto G_{4\phi}$ which appears favoured by observational data \cite{Noller:2018wyv}.

Recalling that, in a homogeneous and isotropic background, the matter stress-energy tensor is $T^{\rm(m)}_{ab}=\rho^{\rm(m)}u_{a}u_{b}+P^{\rm(m)}h_{ab}\,$,
 the pressure in the effective $\phi$-fluid~\eqref{P_cr} is comprised of three contributions, 
\begin{equation}
    P^{(\phi)}= P_{\rm int}+P_{\rm non-visc}+P_{\rm visc},
\end{equation}
the interaction\footnote{The non-minimal coupling of the scalar field with the metric tensor can be translated into an interaction contribution between standard matter and scalar field at the level of field equations.}, non-viscous and viscous pressure, respectively. The viscous pressure, similarly to the case in \cite{Giardino:2022sdv}, is proportional to $H$.\footnote{{ For a detailed discussion on the splitting of viscous and non-viscous terms within this thermodynamic analogy for maximally symmetric spaces we refer the reader to \cite{Giardino:2022sdv}.}} Taking into account Eq.~\eqref{G3}, the explicit expressions of the pressures are
\begin{equation}
      P_{\rm int}=\frac{ \left(G_{4\phi}-F\right)}{G_{4}\Delta}   \left[G_{4\phi}\left(\rho^{\rm(m)}-3 P^{\rm(m)}\right)- 3 F \left(\rho^{\rm(m)}-P^{\rm(m)}\right)\right],
\end{equation}
\begin{align}
      P_{\rm non-visc}=\frac{ 1 }{G_{4}\Delta}&\{2  X   G_{4}  \left(G_{2X} +2  X  G_{2XX} \right) \left(2 G_{4\phi\phi}-F_{\phi}\ln( X/X_{*})\right)\nn\\
    &+G_{2}  \left[ G_{4}  \left(G_{2X} +2  X  G_{2XX} \right)-2  G_{4}  (1+\ln( X/X_{*})) F_{\phi} +4 F   G_{4\phi} -3 F ^2- G_{4\phi} ^2\right]\nn\\
    &+2  X  G_{2X}  \left(3 F ^2-4 F   G_{4\phi} + G_{4\phi} ^2\right)-2  G_{4}  \left(G_{2\phi} -2  X  G_{2\phi X} \right) \left(F - G_{4\phi} \right)\nn\\
    &-2  X  \left[\ln( X/X_{*}) \left(2  G_{4}  F_{\phi\phi}  \left(F - G_{4\phi} \right)+F_{\phi}  \left(-2  G_{4}  \left(F_{\phi} -2 G_{4\phi\phi} \right)-4 F   G_{4\phi} +3 F ^2+ G_{4\phi} ^2\right)\right)\right.\nn\\
    &\left.+4  G_{4}  F_{\phi}  G_{4\phi\phi} -2  G_{4}  \ln^2(X/X_{*}) F_{\phi} ^2\right]\}\,,
\end{align}
\begin{align}
     P_{\rm visc}=-\frac{\epsilon\sqrt{2X}H}{G_{4}\Delta}&\left\{G_{4\phi}  \left[G_{4} \left(G_{2X} -4  X     G_{2XX} +2 (5-\ln( X/X_{*})) F_{\phi} \right)+21 F ^2\right]\right.\nn\\
    &\left.-3 F  G_{4}  \left[2 (1-\ln( X/X_{*})) F_{\phi} +G_{2X} \right]-3 \left(5 F   G_{4\phi} ^2+3 F ^3- G_{4\phi} ^3\right)\right\},\label{Pvisc}
\end{align}
where
\begin{equation}
    \Delta=G_{4} \left[G_{2X}+2 X G_{2XX}-2 (1+\ln( X/X_{*})) F_{\phi}\right]+3 \left(F-G_{4\phi}\right)^2.
\end{equation}

From the viscous component $P_{\rm visc}$ of the pressure, we can extract the bulk viscosity coefficient $\zeta$ as defined in Eq.~\eqref{EckartPvis}, which is proportional to $\dot{\phi}$ similarly to scalar-tensor theories in \cite{Giardino:2022sdv} and reads
\begin{align}\label{bulk}
    \zeta=&\dfrac{\epsilon\sqrt{{2X}}}{3G_4 \Delta}\left\{G_{4\phi}  \left[G_{4} \left(G_{2X} -4  X     G_{2XX} +2 (5-\ln( X/X_{*})) F_{\phi} \right)+21 F ^2\right]\right.\nn\\
    &\left.-3 F  G_{4}  \left[2 (1-\ln( X/X_{*})) F_{\phi} +G_{2X} \right]-3 \left(5 F   G_{4\phi} ^2+3 F ^3- G_{4\phi} ^3\right)\right\}.
\end{align}
Focusing only on dynamical scalar fields, we can have either a vanishing bulk viscosity, 
corresponding to
\begin{equation}
    \begin{split}
    G_{2}(\phi,X)&=\mu(\phi )\left(\frac{5G_{4\phi}(\phi)-3F(\phi)}{4G_{4\phi}(\phi)}\right)^{-1}X^{\frac{5G_{4\phi}(\phi)-3F(\phi)}{4G_{4\phi}(\phi)}}+\nu(\phi)\\
    & \qquad -4X \left( F_{\phi}(\phi )+\frac{3 \left[F(\phi )-G_{4\phi}(\phi)\right]^2}{4G_{4}(\phi )}\right)+2 F_{\phi}(\phi)X\ln( X/X_{*})\,,
    \end{split}
\end{equation}
or a vanishing interaction term, associated with
$F(\phi)=G_{4\phi}(\phi)\,$.
If one imposes both vanishing $P_{\rm int}$ and $P_{\rm visc}$, the scalar field becomes non-dynamical.

Following the argument in \cite{Giardino:2022sdv}, we can still find the ${\cal KT}$ of Horndeski gravity in FLRW, despite the fact that the heat flux density $q_a$ vanishes identically due to homogeneity. Indeed, the general expression for ${\cal KT}$~\eqref{kt} is found in \cite{Giusti:2021sku} for Horndeski theories without specifying to particular geometries. 
Then, substituting $XG_{3X}=F$ from Eq.~\eqref{G3} into Eq.~\eqref{kt}, we find
\begin{equation}
    \mathcal{K} \mathcal{T} =
    %\frac{\dot{\phi}(X G_{3 X}-G_{4 \phi})}{G_{4}}= 
    \epsilon\sqrt{2 X}\frac{ (G_{4 \phi} - F)}{G_{4}}\,.
    \label{ktnew}
\end{equation}
The above quantity is strictly related to the {\it braiding} that measures the strength of kinetic mixing between tensor and scalar perturbations~\cite{Bellini:2014fua, Bettoni:2015wta}. We notice that the relationship $\zeta={\cal KT}/3$, valid for ``traditional'' scalar-tensor theories \cite{Giardino:2022sdv} is not valid for Horndeski theories.

However, Eq.~\eqref{ktnew} leads to an intriguing observation: ${\cal KT}=0$ both for $\dot{\phi}=0$ (which is the usual GR equilibrium) and $F=G_{4\phi}$. The latter is a novel feature of Horndeski gravity in first-order thermodynamics that went unnoticed in \cite{Giusti:2021sku}. It is interesting because it means there are equilibrium states at ${\cal KT}=0$ in the theory that are different than GR. In general, such alternative equilibrium states are found to be unstable \cite{Giardino:2023qlu} and are therefore unable to compete with the special role of GR in the landscape of gravity theories seen through the lens of the first-order thermodynamics. The stability of such states is assessed (generally after reducing to an exact solution of the theory) through the effective heat equation that provides the precise description of the dissipative process leading from non-equilibrium to equilibrium. For Horndeski theories this equation reads \cite{Giusti:2021sku}
\begin{equation}
    \dfrac{\rm d({\cal KT})}{d\tau}=\left(\epsilon\dfrac{\Box\phi}{\sqrt{2X}}-\Theta\right){\cal KT}+\nabla^c\phi\nabla_c\left(\dfrac{G_{4\phi}-XG_{3X}}{G_4}\right),
\end{equation}
where $\dfrac{d}{d\tau}\equiv u^{a}\nabla_{a}=\epsilon\dfrac{\nabla^{a}\phi}{\sqrt{2X}}\nabla_{a}$.

\section{Exact solutions}
\label{sec:4}

In order to test the thermodynamic formulation detailed in the previous sections, we now turn
to studying some exact FLRW solutions. In particular, we focus on background cosmologies in cubic shift-symmetric Horndeski theories with a vanishing scalar current. Since Galileons possess shift symmetry (in addition to Galileian symmetry), this class of theories has some of the most interesting and well-explored cosmological consequences, as mentioned in section \ref{sec:1}. We start in subsection \ref{1st} from the shift-symmetric solution and follow the strategy in \cite{Bernardo:2019vln} to find a cosmological solution with the desired expansion behaviour. From this, we are able to obtain a new, shift-symmetric-inspired solution with explicit scalar field dependence in subsection \ref{2nd}. This can be interpreted as a theory that asymptotically approaches its shift-symmetric formulation.

\subsection{Shift-symmetric gravity}
\label{1st}
The shift symmetry refers to the theory being invariant under $\phi\to\phi+\phi_0$, where $\phi_0$ is a constant. The shift-symmetric subclass of the Horndeski theory corresponds to the choice $G_{i}=G_{i}(X)$, {\it i.e.} the Lagrangian does not explicitly depend on $\phi$. In this case, the theory is characterized by the presence of a Noether conserved current, $J^{a}$, and the scalar field equation of motion assumes the form of $\nabla_{a}J^{a}=0$. 
The shift-symmetric viable Horndeski scalar current is
\begin{equation}
    J^a = \left(  G_{3X} \Box \phi -G_{2X} \right)\nabla^a\phi + G_{3X}\nabla^a X \,.
\label{eq:scalar_current1}
\end{equation}
In the spatially flat FLRW, the scalar current has only the time component, and it reads as follows
\begin{equation}
    J^{a}=\delta^{a}_{0}\dot{\phi}\left(G_{2X}+3HG_{3X}\dot{\phi}\right).
\end{equation}

{
If we restrict the class of ``viable'' shift-symmetric Horndeski theories, we have that the shift-symmetry sets $G_{i}=G_{i}(X)$ while the ``viability'' requires the conditions $G_{4X}=G_5=0$. Combining the two, one finds that $G_4 = \mbox{constant}$ and all non-minimal couplings disappear. Then, taking the covariant divergence of Eq.~\eqref{Einstein}, recalling Eq.~\eqref{teff} and the contracted Bianchi identity, one finds
$$
\nabla^a T^{( \phi)}_{ab} = G_4^{-1} \, \nabla^a T^{(\rm m)}_{ab} \, .
$$
Thus, at the level of the field equations for the metric tensor, the conservation of the stress-energy tensor of matter implies that of $T^{( \phi)}_{ab}$ in this specific scenario. 
Another way of understanding this point consists in observing that for this class of models $\nabla^{a}T^{( \phi)}_{ab}=-\tfrac{1}{2}(\delta L/\delta\phi)\nabla_{b}\phi \big|_{\rm on-shell} = 0$. Hence, since $G_4 = \mbox{constant}$ and $T^{({\rm eff})}_{ab} = G_4^{-1} \, T^{(\rm m)}_{ab} + T^{( \phi)}_{ab}$, then $\nabla^{a}T^{\rm(eff)}_{ab}=0$ comes from the independent covariant conservation of both $T^{(\rm m)}_{ab}$ and $T^{( \phi)}_{ab}$.}

The following gravitational action describes the shift-symmetric sector of the linear model selected in the previous section
\begin{equation}
\label{shift}
S_{g} = \frac{1}{2}\int d^4 x \sqrt{-g} \left[ \,R + G_2(X) - \lambda\ln{(X/X_{*})} \Box \phi \, \right],
\end{equation}
so that $G_2=G_2(X)$, $G_3 = \lambda \ln{(X/X_{*})}$, and $G_4=1$. 
The study of this choice of couplings is also motivated from a phenomenological point of view, since it provides a good fit to cosmological data from standard probes \cite{Arjona:2019rfn, Noller:2018wyv}.

Then, the associated scalar current reduces to
\begin{equation}
    J^a = \delta^{a}_0\left(\dot{\phi}\,G_{2X}+6\lambda H\right).
\label{eq:scalar_current2}
\end{equation}
In this work, we just restrict to the solutions
associated with vanishing scalar current, in order to provide some concrete examples, similarly to \cite{Bernardo:2019vln}. Excluding the trivial case of $\dot{\phi}=0$ which is equivalent to GR, the vanishing scalar current entails
\begin{equation}
    \dot{\phi}\,G_{2X}+6\lambda H=0\,.\label{ourshiftcurrent}
\end{equation}

Using the above equation, and assuming that the standard matter content is described by the linear barotropic equation of state $P^{(m)}=w\rho^{(m)}$, we obtain the following expressions for the scalar field energy density and pressure, respectively,
\begin{eqnarray}
     \rho^{(\phi)}&=&-\frac{1}{2}G_2\,,\label{rho_shift}\\
    &&\nn\\
      P^{(\phi)}&=&\frac{1}{2}G_2-\frac{ X  G_{2X} ^2+3 \lambda^2 G_{2} }{G_{2X} +2  X  G_{2XX} +3 \lambda^2}-\frac{3  \lambda^2 (w-1) \rho^{\rm(m)}}{G_{2X} +2  X  G_{2XX} +3 \lambda^2}\,.\label{P_shift}
\end{eqnarray}

The effective scalar fluid temperature of shift-symmetric viable Horndeski then reads
\begin{equation}\label{kt_shift}
    \mathcal{KT}=-\lambda\epsilon\sqrt{2X}\,.
\end{equation}
If we want a positive defined $\mathcal{KT}$, then $\lambda\epsilon$ must be negative, {\it i.e.}, $\lambda\,{\rm Sign}(\dot{\phi})>0$. Since $X$ must be strictly positive, the above temperature will not reach the zero temperature equilibrium state associated to GR. This shows that the approach to equilibrium is not always granted, as it was also found in \cite{Giardino:2022sdv}.

In order to find exact analytical solutions, we assume $P^{\rm(m)}=w\rho^{\rm(m)}$, $H$ strictly monotonic ({\it i.e.}, we can write $t$ as $t = t(H)$), so that Eq.~\eqref{time_feqs} and the scalar equations of motion~\eqref{ourshiftcurrent} respectively read as follows
\begin{eqnarray}
G_2\left[H\right] &=& 2 \rho^{\rm (m)} \left[H\right] - 6 H^2\,,\label{t_feq}\\
G_{2X}\left[H\right] &=& -\frac{6\lambda H}{ \dot{\phi}\left[H\right]}=\frac{6\lambda\epsilon H }{\sqrt{2X[H]}}\,,\label{vanish_J}
\end{eqnarray}
where $\epsilon=-{\rm Sign}(\dot{\phi})$. Differentiating $G_{2}$ with respect to $H$ one gets
$$
\frac{dG_2}{dH} = G_{2X}\frac{dX}{dH} \, ,
$$
and differentiating Eq.~\eqref{t_feq} with respect to $H$ one obtains
$$
\frac{dG_2}{dH} = 2\frac{d\rho^{\rm(m)}}{dH}-12H \, .
$$
Combining the last two equations yields
$$
2\frac{d\rho^{\rm(m)}}{dH}-12H
=
G_{2X}\frac{dX}{dH} \, .
$$
Then, taking advantage of Eq.~\eqref{vanish_J}, the latter can be rewritten as
$$
2\frac{d\rho^{\rm(m)}}{dH}-12H
=
\left(\frac{6\lambda H}{\epsilon\sqrt{2X}}\right)\frac{dX}{dH} \, .
$$
The last equation can be {\em formally} integrated on both sides in $H$ as
\begin{equation}
 \int{\left(\frac{2}{H}\frac{d\rho^{\rm(m)}}{dH}-12\right)dH}
 = {6\lambda\epsilon}\int{\frac{dX}{\sqrt{2X}}} \, , \label{HX}   
\end{equation}
or, equivalently,
\begin{equation}
 \int{\left(\frac{2}{H}\frac{d\rho^{\rm(m)}}{dH}\right)dH} - 12 H
 = {6\lambda\epsilon} \,\sqrt{2X} \, . \label{HX2}   
\end{equation}

Inspired by the strategy used in \cite{Bernardo:2019vln}, our approach consists of choosing a cosmological evolution, either power-law expansion or exponential expansion, and then solving the continuity equation for the matter perfect fluid energy density, such that we can analytically obtain the function $G_{2X}$ by inverting the relation $X[H]$ (if possible) and integrating the vanishing scalar current condition.

Let us start by considering a power-law expanding universe,
\begin{equation}\label{eq:powlaw1}
    a(t)=a_{*}\left(\frac{t}{t_{*}}\right)^{n},\quad n>0\,,\enspace t\ge0\,,
\end{equation}
with $a_{*}$ constant.
As a consequence, the following equations hold
\begin{eqnarray}
    H(t)=\frac{n}{t}\quad&\longleftrightarrow&\quad t(H)=\frac{n}{H}\,,\label{eq:powlaw2}\\
    \rho^{\rm(m)}(t)=\rho_{*}\left(\frac{t_{*}}{t}\right)^{3n(w+1)}\quad&\longleftrightarrow&\quad \rho^{\rm(m)}(H)=\rho_{*}\left(\frac{H}{H_{*}}\right)^{3n(w+1)},\label{eq:powlaw3}
\end{eqnarray}
where all the $*$-quantities are constant. Using Eqs.~\eqref{eq:powlaw2} and~\eqref{eq:powlaw3} into Eq.~\eqref{HX2}, and performing the integration, we obtain
\begin{equation}
    6\lambda\epsilon\sqrt{2X}=6\lambda\epsilon\,c_{X}-12 H+\frac{6  n \rho_{*} (w+1) }{H_{*} [3 n (w+1)-1]}\left(\frac{H}{H_{*}}\right)^{3 n (w+1)-1},\label{eq:powlaw4}
\end{equation}
where $c_X$ is an integration constant, and it is associated with the non-vanishing asymptotic value of $\sqrt{2X}$ reached in correspondence with $H=0$ (in the limit $t\to\infty$). 

The simplest case admitting an analytical solution corresponds to $n=2/3(w+1)$, with $w\neq-1$\,. Then, it turns out that
\begin{equation}\label{eq:sqrtx1}
    6\lambda\epsilon\sqrt{2X}=12 H \left(\frac{  \rho_{*}}{3H_{*}^2}-1\right)+6\lambda\epsilon\,c_X\quad\Rightarrow\quad H=\frac{\lambda\epsilon}{2}\left(\sqrt{2X}-c_X\right)\left(\frac{  \rho_{*}}{3H_{*}^2}-1\right)^{-1}\,.
\end{equation}
Thus the system is analytically solvable, yielding
\begin{equation}\label{G2_shift}
    G_{2}(X)= \frac{3\lambda^2}{2}\left(\sqrt{2X}-c_X\right)^{2} \left(\frac{  \rho_{*}}{3H_{*}^2}-1\right)^{-1}\,,
\end{equation}
with the scalar field having the following form
\begin{equation}
    \phi (t)=\phi_{*} -\epsilon\, c_{X}( t-t_{*})  -\frac{2n}{3\lambda} \left( \frac{  \rho_{*}}{3 H_{*}^2}-1\right) \ln(t/t_{*})\,,
\end{equation}
where $\phi_{*}$ is constant.

Therefore, the effective scalar fluid temperature reads
\begin{equation}
    \mathcal{KT}= \frac{2n}{t} \left(1-\frac{  \rho_{*}}{3H_{*}^2}\right)-\lambda\epsilon\,c_X\,.
\end{equation}
As expected for cosmological solutions with an initial singularity \cite{Giardino:2022sdv}, ${\cal KT}\rightarrow+\infty$ for $t\rightarrow 0$, indicating an extreme deviation of Horndeski theory from the GR equilibrium state as the singularity is approached. 

We can now make a sensible consideration about the sign of the constants appearing in our solution. First, let us take into account the case $\mathcal{KT}>0$. This implies $\lambda\epsilon<0$ because of Eq.~\eqref{kt_shift}. However, there is an additional condition to be satisfied in order to have a positive definite $\mathcal{KT}$ for any $t>0$, namely $\left(1-\dfrac{ \rho_{*}}{3H_{*}^2}\right)(w+1)>0$, which is associated with the requirement of a non-vanishing kinetic term. Then, assuming $w>-1$ (corresponding to $n>0$), we obtain $\left(1-\dfrac{ \rho_{*}}{3H_{*}^2}\right)>0$. Therefore, the term $\sqrt{2X}$ starts from an infinite positive value, corresponding to the initial singularity at $t=0$, and approaches $c_X$ for $t\to\infty$. This is consistent with Eqs.~\eqref{eq:powlaw2} and~\eqref{eq:sqrtx1}, where $H=\dfrac{n}{t}$ must be positive. As a consequence, $G_2$ is actually negative definite, and this implies $\rho^{(\phi)}>0$ from~\eqref{rho_shift}. It is straightforward to verify that $\rho^{(\phi)}=\dfrac{3 n^2}{t^2}\left(1-\dfrac{ \rho_{*}}{3H_{*}^2}\right)$ and $P^{(\phi)}=w \rho^{(\phi)}$. Therefore, $\rho^{(\phi)}+P^{(\phi)}=\rho^{(\phi)}(1+w)>0$ is satisfied. In this sense, the definition of effective temperature obtained by generalizing the one found for “traditional” scalar-tensor first-order thermodynamics implies the weak energy condition for the effective fluid. It is interesting to notice that this consideration is independent of ${\rm Sign}(\dot{\phi})$, which is involved only in the condition $\lambda\epsilon<0$.
Lastly, the bulk viscosity coefficient $\zeta$ (cfr. Eq.~\eqref{bulk}) yields
\begin{equation}
    \zeta=\lambda  \epsilon  \left(\sqrt{2X}-c_X\frac{3H_{*}^2}{  \rho_{*}}\right) =   - \frac{2n}{t} \left(1-\frac{  \rho_{*}}{3H_{*}^2}\right)+\lambda\epsilon c_X\left(1-\frac{3H_{*}^2}{  \rho_{*}}\right)\,.
\end{equation} 
The above equation shows that the effective fluid starts off with a negative (and diverging to $-\infty$) bulk viscosity approaching the initial singularity, then $\zeta$ vanishes as the gradient of the scalar field approaches $\sqrt{2X}=c_X \dfrac{3H_{*}^2}{  \rho_{*}}$, or, equivalently, as the cosmological time approaches $t= -\dfrac{1}{\lambda\epsilon}\dfrac{2 n \rho_{*}}{3c_X H_{*}^2 }$, and finally it becomes positive as $t$ increases from that point.\\

Let us now take into account the case of a spatially flat de-Sitter spacetime,
\begin{equation}
    a(t)=a_{*}\exp(H_{*}t)\,.
\end{equation}
The continuity equation gives
\begin{equation}
    \rho^{\rm(m)} (t)= \rho_{*} \exp{[-3H_{*} (w+1)t]},
\end{equation}
and, from Eq.~\eqref{vanish_J}, we obtain
\begin{equation}
    G_{2}(X)=6 \lambda H_{*} \left(\epsilon\sqrt{2X}+c_X\right).
\end{equation}
Then, from the temporal component of the field equations~\eqref{t_feq}, we obtain
\begin{equation}
    6 \lambda H_{*}( \epsilon \sqrt{2X}+c_X)=2 \rho^{\rm(m)} - 6 H_{*}^2,
\end{equation}
which is equivalent to
\begin{equation}
    \dot{\phi}=c_{X}+\frac{H_{*}}{\lambda }\left(1-\frac{\rho_{*} }{3H_{*}^2}\exp{[-3 H_{*} (w+1)t]}\right).
\end{equation}
Integrating the equation above, the scalar field reads
\begin{equation}
    \phi(t)=\phi_{*}+t \left(c_{X}+\frac{H_{*}}{\lambda }\right)+\frac{1}{3 (w+1)\lambda}\frac{  \rho_{*} }{3 H_{*}^2 }\exp{[-3 H_{*}  (w+1)t]}\,.
\end{equation}
The effective scalar fluid temperature is
\begin{equation}
    \mathcal{KT}= (\lambda\epsilon\,c_{X}+H_{*}) -\frac{  \rho_{*} }{3 H_{*}}\exp{[-3 H_{*}  (w+1)t]}.
\end{equation}
Also in this case, the condition $\left(1-\dfrac{  \rho_{*}}{3H_{*}^2}\right)>0$ ensures the positivity of $\rho^{(\phi)}$ and of $\mathcal{KT}$, under the assumption of $\epsilon\lambda<0$ and $c_{X}<-\dfrac{H_{*}}{\epsilon\lambda}\left(1-\dfrac{ \rho_{*}}{3H_{*}^2}\right)$. The constant $c_X$ can be properly chosen so that the condition $\rho^{(\phi)}+P^{(\phi)}>0$ is satisfied as well. Therefore, the effective fluid can be easily tuned to satisfy the weak energy condition, characteristic of a real fluid. Also in this case, imposing the positivity of $\mathcal{KT}$ goes in the direction of recovering the weak energy condition.

\subsection{New exact solution with asymptotic shift-symmetry}
%New exact asymptotically shift-symmetric solution
\label{2nd}
Using a heuristic approach, we can generalize the previous power-law solution, adding an explicit $\phi$-dependence inside the action. In particular, we consider the case described by Eqs.~\eqref{eq:powlaw1}-\eqref{eq:powlaw3}, within $n=2/3(w+1)$, and, inspired by Eq.~\eqref{eq:sqrtx1}, we assume the following equation
\begin{equation}\label{eq:sqrtx2}
    H(t)=\alpha\sqrt{2X}+\beta\,,
\end{equation}
where, $\alpha$ and $\beta$ are constants. The expression above provides a differential equation for the scalar field, implying 
\begin{equation}\label{eq:s}
    \dot{H}(t)=-\alpha\epsilon\,\ddot{\phi}(t)\,.
\end{equation}
Since we already have fixed the scalar field time-dependence, once we use Eqs.~\eqref{eq:sqrtx2} and~\eqref{eq:s}, we must require the scalar field equation of motion to be identically solved by the functional form of the action,
\begin{equation}
\label{gen-shift}
S_{g} = \frac{1}{2}\int d^4 x \sqrt{-g} \left[ \,G_{4}(\phi)R + G_2(\phi,X) - F(\phi)\ln{(X/X_{*})} \Box \phi \, \right].
\end{equation}
Therefore, let us write down the equation of motion of the scalar field
\begin{eqnarray}
    \ddot{\phi} \left[\,2F_{\phi} \ln{(X/X_{*})} +2 F_{\phi}+6 \alpha  \epsilon  \left(F-G_{4\phi}\right)-G_{2X} -2 X G_{XX}\,\right]&&\nn\\
    +X \left[6 \alpha  \left(\epsilon  G_{2X} +2 \epsilon   F_{\phi} -6 \alpha  F +4 \alpha   G_{4\phi} \right)-2 G_{2\phi X} \right]&&\nn\\
    +\ln( X/X_{*}) \left[2 X \left(F_{\phi\phi}-6 \alpha  \epsilon   F_{\phi} \right)-6  \beta  \epsilon  \sqrt{2X}  F_{\phi} \right]&&\nn\\
    +3\beta  \sqrt{2X} \left[\epsilon  G_{2X} +2 \epsilon   F_{\phi} -12 \alpha  F +8 \alpha   G_{4\phi} \right]&&\nn\\
    +G_{2\phi} +6 \beta ^2 \left[2  G_{4\phi} -3 F \right]&&=0\,.
\end{eqnarray}
First, we need to impose that the coefficient of $\ddot{\phi}$ vanishes, {\em i.e.},
\begin{equation}
    2F_{\phi} \ln{(X/X_{*})} +2 F_{\phi}+6 \alpha  \epsilon  \left(F-G_{4\phi}\right)-G_{2X} -2 X G_{XX}=0\,,
\end{equation}
which can be seen as a differential equation for $G_2$, having as solution the following functional form,
\begin{equation}\label{G2}
    G_{2}(\phi,X)= \mu\sqrt{2X} +\nu+2 X F_{\phi}  \ln (X/X_{*}) -X   \left[4 F_{\phi}-6 \alpha  \epsilon \left( F-G_{4\phi}\right) \right]\,,
\end{equation}
where $\mu=\mu(\phi)$ and $\nu=\nu(\phi)$ are integrating functions of the scalar field. Substituting Eq.~\eqref{G2} into the field equations of the scalar field and of the metric tensor yields
\begin{align}
    &  6\alpha\left[  \epsilon  \left(G_{4\phi\phi}-F_{\phi} \right)-2 \alpha  G_{4\phi}\right]\,X+ 3\alpha\left[2  \beta  \left( G_{4\phi}-3 F \right)+   \epsilon  \mu \right]\sqrt{2X }+6\beta ^2 \left(2 G_{4\phi}-3 F \right)+3 \beta  \epsilon  \mu +\nu_{\phi}=0\,,\\
    &\nn\\
    &  3\left[  \epsilon  \left( F - G_{4\phi}\right)+2\alpha  \left( G_{4}-\frac{  \rho_{*}}{3H_{*}^2}\right)\right]\left(\alpha X+ \beta\sqrt{2X}\right) +3\beta ^2 \left( G_{4}-\frac{   \rho_{*}}{3H_{*}^2}\right)+\frac{1}{2} \nu =0.
\end{align}
The above equations appear to have the same structure with respect to the $X$-dependence, namely $c_1(\phi) X+c_2(\phi)\sqrt{2X}+c_{3}(\phi)=0$.  
We can obtain the general solution associated with Eqs.~\eqref{eq:sqrtx2} and~\eqref{eq:s}, by requiring all coefficients $c_i(\phi)$ of the above equations to vanish. This provides the following (unknown) functions of the scalar field,
\begin{align}
    F(\phi)=&\,2\alpha\epsilon\left(\frac{ \rho_{*}}{3H^2_{*}}-G_{4}\right)+G_{4\phi}\,,\\
    \mu(\phi)=&\,4\beta\epsilon\left[3\alpha\epsilon\left(\frac{ \rho_{*}}{3H^2_{*}}-G_{4}\right)+G_{4\phi}\right],\\
    \nu(\phi)=&\,6 \beta ^2 \left(\frac{   \rho_{*}}{3H_{*}^2}-G_4\right),
\end{align}
while the effective temperature reads
\begin{align}
    \mathcal{KT}=2 \alpha \sqrt{2X}  \left(1-\frac{1}{G_4}\frac{   \rho_{*}}{3H_{*}^2}\right).
\end{align}
Then, the only remaining free Horndeski function is $G_{4}$. When the non-minimal coupling function is equal to one, the action~\eqref{gen-shift} reduces to that of the shift-symmetric theory, provided that
\begin{equation}
    \alpha =\frac{\lambda  \epsilon}{2} \left(\frac{   \rho_{*}}{3H_{*}^2}-1\right)^{-1},\qquad\beta=-\frac{\lambda  \epsilon}{2} c_X \left(\frac{   \rho_{*}}{3H_{*}^2}-1\right)^{-1}.
\end{equation}
It is useful to understand the solution considered above as asymptotically approaching its shift-symmetric formulation. In this case, the remark about the positivity of $\mathcal{KT}$ translates to $G_{4}>\dfrac{ \rho_{*}}{3H_{*}^2}$. This means that $G_{4}$ cannot be chosen such that the initial singularity is removed and the original singularity of the shift-symmetric model remains unchanged.

\section{Conclusions}

In this work, we have specialised the formalism for the first-order thermodynamics of viable Horndeski gravity to the case of spatially flat isotropic and homogeneous spacetimes, with the goal of testing the physical intuition behind the formalism in this class of theories with interesting implications. 

In general, the thermodynamics of scalar-tensor gravity \cite{Faraoni:2021jri, Faraoni:2021lfc} relies on the {\em effective fluid approach} to modified gravity theories \cite{Pujolas:2011he, Faraoni:2018qdr}. According to this framework, given an alternative theory of gravity, we can derive from its field equations a generalised Einstein equation where all the contributions of the additional scalar degree of freedom -- other than the matter fields -- are collected in an effective stress-energy tensor $T^{(\phi)}_{ab}$. Such an effective tensor is symmetric by construction: thus, given a timelike vector field $u^a$, $T^{(\phi)} _{ab}$ always admits an imperfect fluid decomposition based on $u^a$ (this is a trivial algebraic result, though for details we refer the reader to \cite{Faraoni:2023hwu}). In other words, we can always find effective energy density, pressure, heat fluxes, and anisotropic stresses associated to $T^{(\phi)} _{ab}$ and $u^a$. More importantly, if $\nabla^a \phi$ is timelike, we can construct a 4-velocity field $u^a \propto \nabla^a \phi$, allowing us to provide a fluid interpretation for $T^{(\phi)} _{ab}$. This fluid, with 4-velocity $u^a \propto \nabla^a \phi$, is the dubbed $\phi$-fluid. Studying the kinematic quantities of such an effective fluid and comparing them to the imperfect fluid decomposition of $T^{(\phi)} _{ab}$ one can then infer the constitutive relations of the $\phi$-fluid. It was then shown in \cite{Giusti:2021sku}, generalised in \cite{Miranda:2022wkz}, and further expanded on in this work, that the effective fluid representation for the viable subclass of Horndeski gravity satisfies the constitutive laws of Eckart's theory of non-equilibrium thermodynamics. This allows to define an effective temperature for the $\phi$-fluid, which is positive-definite for scalar-tensor theories and represents the order parameter characterising the approach (or lack thereof) to the GR equilibrium state at zero temperature.

Since Eckart's constitutive relations are linear in the velocity gradient $\nabla_{b} u_{a}$, specialising the analysis for cosmological backgrounds further restricts the Horndeski theory to the subclass characterised by $G_{3}=F(\phi)\ln(X/X_{*})$. 

Contrary to ``traditional'' scalar-tensor theories \cite{Giardino:2022sdv}, viable Horndeski gravity naturally admits zero temperature equilibrium states other than GR (which is characterised by $\phi =$~const.) corresponding to the condition $\sqrt{2X}(G_{4\phi}-XG_{3X})=0$, {\it i.e.} $G_{3}=G_{4}\ln(X/X_{*})$. This class of zero-temperature equilibrium states alternative to GR is characterised by non-vanishing viscosity coefficients, thus suggesting that such equilibrium states are actually {\em unstable}.  

In flat FLRW cosmology, due to the symmetries of the background, the heat flux and the anisotropic stress vanish identically. However, the viscous contribution remains and is visible through the isotropic pressure giving rise to a non-vanishing bulk viscosity. We computed the effective bulk viscosity for such models in this scenario, while the temperature and thermal conductivity are naturally inherited from the general (background-independent) approach. 

The general results for the thermodynamics of viable Horndeski cosmology were then tested against exact solutions for interesting subclasses of the general theory that are also favoured by cosmological observations. The considered examples differ significantly from the results obtained for ``traditional'' scalar-tensor cosmologies, since they display a non-vanishing effective temperature at all times in the cosmic evolution and asymptotically approach a constant effective temperature at late times. These results have been obtained, in particular, for classes of shift-symmetric and asymptotically shift-symmetric theories (the latter being shift-symmetric as the non-minimal coupling function $G_4$ approaches unity), both characterised by a non-vanishing braiding parameter. 

In addition to showing the existence of subclasses of viable Horndeski gravity that never relax to the GR equilibrium state, our analysis further confirms previous findings according to which curvature singularities are ``hot'' \cite{Faraoni:2021jri}, exhibiting a diverging temperature. This suggests that the deviations of these models from General Relativity become extreme at spacetime singularities. An additional intriguing consequence of finding the effective temperature associated to these viable Horndeski subclasses is that imposing its positivity recovers the weak energy condition for the $\phi$-fluid, which is characteristic of a real fluid and was not expected to hold for an effective fluid.

Lastly, in this work we have also provided a novel exact cosmological solution for an asymptotically shift-symmetric theory as a toy model for our thermodynamic analysis. 
\begin{acknowledgments} 
The authors would like to thank Valerio Faraoni and Nicola Muttoni for useful discussions. M.M. is grateful for the support of Istituto Nazionale
di Fisica Nucleare (INFN) iniziativa specifica MOONLIGHT2. The work of A.G. and S.G. has been carried out in the framework of the activities of the Italian National Group of Mathematical Physics [Gruppo Nazionale per la Fisica Matematica (GNFM), Istituto Nazionale di Alta Matematica (INdAM)]. L.H. is supported by funding from the European Research Council (ERC) under the European Unions Horizon 2020 research and innovation programme grant agreement No 801781. LH further acknowledges support
from the Deutsche Forschungsgemeinschaft (DFG, German Research Foundation) under Germany’s Excellence Strategy EXC 2181/1 - 390900948 (the Heidelberg STRUCTURES Excellence Cluster).
\end{acknowledgments}

%%%%%%%%%%%%%%%%%%%%%%%%%%%%%%%%%%%%%%%%%%%%%%%%%%%%%%%%%
\bibliographystyle{apsrev4-1}
\bibliography{ref}{}

\end{document}